\documentstyle[aps]{revtex}
\newcommand{\mybox}{\hfill{\scriptsize\fbox{\hspace*{1.1pt}}}}
\newcommand{\beq}{\begin{equation}}
\newcommand{\beqa}{\begin{eqnarray}} 
\newcommand{\eeqa}{\end{eqnarray}} 
\newcommand{\eeq}{\end{equation}}
\def\reals{\hbox{\rm I\kern -.2em R}}

\def\com{\hbox{\rm l\kern -.4em C}}

\def\rhom{\rho_{M}}
\def\orho{{\cal O}_{g}(\rho)}
\def\pt{\mbox{pt}}
\def\ctr{^{\dagger}}
\def\rob{\mbox{R}_{g}(\rho)}

\def\sep{{\cal S}}
\def\tt{{\cal T}}
\def\dm{\mbox{DM}}
\def\rb{\mbox{R}_{g}}
\def\te{\tilde{e}}
\def\ati{\tilde{a}_{i}}
\def\atj{\tilde{a}_{j}}

\def\ulr{|r>}

\def\uls{|s>}
\def\ulsp{<s|}
\def\ult{|t>}
\def\ultp{<t|}
\def\ulu{|u>}
\def\ulup{<u|}
\def\ulj{|j>}
\def\uljp{<j|}
\def\ulk{|k>}
\def\ulkp{<k|}
\def\dfn{\hbox{\rm $=$\kern -.95em $~^{^{\triangle}}$}}

\newtheorem{theorem}{Theorem}
\newtheorem{lemma}{Lemma}

\begin{document}

\draft

\title{Generalized Robustness of Entanglement}
 
\author{Michael Steiner}
\address{Naval Research Laboratory}
\date{\today}
\maketitle 

\begin{abstract} 
The robustness of entanglement results of Vidal and Tarrach considered the problem whereby an entangled state is mixed with a separable state so that the overall state becomes non-entangled. In general it is known that there are also cases when entangled states are mixed with other entangled states and where the sum is separable. In this paper, we treat the more general case where entangled states can be mixed with any states so that the resulting mixture is unentangled. It is found that entangled pure states for this generalized case have the same robustness as the restricted case of Vidal and Tarrach. \end{abstract}


\section{Introduction}

The  Robustness of Entanglement  in \cite{Q1:Vidal2} examines how much mixing can take place between an entangled state $\rho$ and another state $\rho_{M}$, so that the convex combination of the two states is separable. Such work has significance to how robust entangled states are in the presense of interaction mechanisms that could disentangle the state.  This work is also of interest in the characterization of the state space in terms of entangled and separable states, and the decomposition is an important result in quantum information theory. 

In \cite{Q1:Vidal2} the authors restricted $\rho_{M}$ to be separable. This is a reasonable restriction as long as the states that are being mixed are separable.  Another possibility is that a given entangled state would not only interact with separable states, but could also interact with other entangled states. It is known that the mixing of entangled states can also result in the convex combination being separable.  Hence entangled states thought to be robust for the cases where interaction is only with separable states, might not be robust when allowed to interact with arbitary states.  This would require the presense of entangled states in the interaction medium that are stable at least within the interaction time.  The degree and scale for which  entangled states exist is not yet known and is a current area of research in mesoscopic physics.  What is known is that the processes of entanglement and decoherence are continually at work and it becomes more difficult to maintain entanglement with the scale of the entangled state. In general it is known that microscopic entangled states are found that are very stable, for example electron-sharing in atomic bonding and two-particle entangled photon states generated by parametric down conversion.  Additionally, it is known that certain larger entangled systems can exist. Examples of multi-particle superpositions that are given in \cite{Q1:Zeh2} include phonons in solids, superfluids, and superconducting quantum interference devices.  It is also known that Bose-Einstein Condensates are examples of large scale superpositions and the relation to entanglement is explored in \cite{Q1:Sorensen}. Hence it is known that entangled states exist on the microscopic level and under certain conditions on the mesoscopic level.

Therefore, the restriction that $\rho_{M}$ be separable is lifted so that $\rho_{M}$ can be an arbitary density matrix. As it is known \cite{Q1:Zyczkowski} that the relative volume of the state space is dominated by entangled states as the dimension of the composite Hilbert space grows, one might initially expect that entangled states would not be as robust when entangled states are allowed to interact with arbitrary states versus only separable states.  Numerical simulations in \cite{Q1:Zyczkowski} indicate that the probability of finding a non-entangled state decreases exponentially with the size of the Hilbert space of the composite system. Geometrically, the separable states become sandwiched between two hyperplanes \cite{Q1:Steiner4}.  However, as we will see, the robustness of entanglement of pure entangled states does not change when $\rho_{M}$ can be an arbitrary state, compared to the case when $\rho_{M}$ is separable. That is, we find the same expression for the robustness of entanglement for the two cases.

\section{Generalized Robustness of Entanglement}

Consider two systems of particles $p_{1},p_{2}$ with composite states represented by density matrices $\rho$, $\rho_{M}$ that operate on the Hilbert space ${\cal C}^{n}\otimes {\cal C}^{n}$, $N=n^{2}$. For a given entangled state $\rho$, Vidal and Tarrach \cite{Q1:Vidal2} considered the problem of finding the largest $a\in[0,1]$ for which there exists a $\rho_{M}\in \sep $ and that
\[ a \rho + (1-a) \rhom \in \sep , \] where $\sep$ denotes the set of separable states. The robustness of entanglement \cite{Q1:Vidal2} was defined as $\mbox{R}_{s}(\rho)\dfn \frac{1}{a}-1$, where $a$ is largest.  Define the optimal $a$ for a given $\rho$ as ${\cal O}_{s}(\rho)$. Note from the definition that for any $a> {\cal O}_{s}(\rho)$, the matrix $a \rho + (1-a) \rho_{M}$ is necessarily entangled for all $\rhom \in \sep$.  For the case where $\rho$ is a pure  state (i.e. rank 1), $\rho=\psi\psi'$ and with $\psi$ having a Schmidt decomposition $\psi=\sum_{i} \tilde{a}_{i} |i> \otimes |i>$, $\tilde{a}_i\ge 0$, it was shown in \cite{Q1:Vidal2} 
that \beq \label{eqnrob} \mbox{R}_{s}(\rho) = (\sum_{i} \tilde{a}_{i})^{2} - 1 .\eeq

For the case of generalized robustness of entanglement, we define  ${\cal O}_{g}(\rho)$ as the largest value of $a$ for which there exists $\rho_{M}\in \dm(N) $ with
\beq \label{eqndef} a \rho + (1-a) \rhom \in \sep , \eeq
where $\dm(N)$ is the set of $N$ by $N$ density matrices. Note that since  ${\cal S}\subset\dm(N)$ then 
\beq \label{eqngenandvt} {\cal O}_{g}(\rho) \ge {\cal O}_{s}(\rho). \eeq
The generalized robustness of entanglement is defined as $\rob\dfn \frac{1}{{\cal O}_{g}(\rho)}-1$.

Several of the results in \cite{Q1:Vidal2} can be extended to the generalized robustness of entanglement case. The proofs of these extensions are straightforward and are shown in the Appendix. We state these results here.

\begin{theorem}
$\rob$ is convex, i.e. \[ \mbox{R}_{g}(t \rho_{1}+ (1-t) \rho_{2}) \le t \mbox{R}_{g}(\rho_{1})+ (1-t) \mbox{R}_{g}(\rho_{2}) .\]
\end{theorem}

\begin{theorem}
\label{thmtwo}
$\rob = \mbox{R}_{g}(U_{L} \rho U_{L}) $, where $U_{L}$ is a local unitary transformation of the form $U_{L}= U_{1} \otimes U_{2}$.
\end{theorem}

\noindent {\bf Corollary} Let $\rho = \psi \psi'$, $\rho$ act on $\com^{n} \otimes \com^{n}$ and $\psi = \sum_{i=1}^{n} \tilde{a_{i}} \gamma_{i}^{A} \otimes \gamma_{i}^{B} $ is a Schmidt decomposition of $\psi$ where $\gamma_{i}^{A}$ is an orthonormal basis for subsystem $A$ and $\gamma_{i}^{B}$ is an orthonormal basis for subsystem $B$. Then if $\tilde{\rho}=\tilde{\psi}\tilde{\psi'}$, $\tilde{\psi} = \sum_{i=1}^{n} \tilde{a_{i}} |i> \otimes |i> $, where $\{|1>, |2>,\cdots,|n>\}$ is the natural basis of $\com^{n}$, then the robustness is the same, i.e. $\mbox{R}_{g}(\tilde{\rho}) = \rob$. 
\vspace{.125in}

\noindent Proof: Define $U_{1}$ by the map $\psi_{i}^{A} \rightarrow |i>$ and similarly define $U_{2}$ for subsystem $B$.  Since change of orthogonal basis mappings are unitary, one can apply Theorem \ref{thmtwo} and the result follows. \mybox

\noindent \begin{lemma}
\label{leig}
Let $\rho$ be a pure state acting on $\com^{n}\otimes \com^{n}$. The eigenvalues of $\rho^{\pt}$ consist of up to $n(n-1)/2$ negative eigenvalues, where $\rho^{\pt}$ denotes the partial transpose \cite{Q1:Peres} of $\rho$. If $\rho=\psi\psi'$,  $\psi = \sum_{i=1}^{n} \tilde{a_{i}} |i> \otimes |i> $, then these eigenvalues are given by $\{ - \tilde{a_{r}} \tilde{a_{s}}\}$, $r<s$. The corresponding eigenvectors are \[ \tilde{e}_{f(r,s)} = 1/\sqrt{2}(|r>|s> - |s>|r>), ~~~r<s\] where $f(i,j)=(j-i)+n(i-1)-i(i-1)/2$, $i<j$.
\end{lemma}

\noindent Proof: The proof is given in \cite{Q1:Vidal2}, Equations B17-B18.

A main result is that for pure states, the generalized robustness of entanglement is the same as the Vidal and Tarrach robustness of entanglement.

\begin{theorem} Let $\rho$ be a pure state. Then ${\cal O}_{g}(\rho)= {\cal O}_{s}(\rho) = \frac{1}{1+R_{s}(\rho)}$. \end{theorem}

\noindent Proof:  Consider the eigenvectors $\te_{i}$ from Lemma \ref{leig}. If $\orho=t$, then there exists a $\rho_{M}\in \mbox{DM(N)}$, with
\beq \label{eqnone} \te_{i}' (t \rho^{\pt} + (1-t) \rho_{M}^{\pt}) \te_{i} \ge 0, ~~~ i=1,\cdots,n(n-1)/2 , \eeq
where $x'$ denotes the conjugate transpose of $x$.
It will be shown that $t\le {\cal O}_{s}(\rho),$ otherwise at least one term in (\ref{eqnone}) will be negative for every $\rho_{M}$. To this end, let $\te'_{f(i,j)} ( t \rho^{\pt}+ (1-t) \rho_{M}^{\pt})\te_{f(i,j)} \ge 0 $ for $j>i$. Then 
\beq \label{eqntwo}  t \le (1+\frac{\ati\atj}{\te_{f(i,j)}'\rho_{M}^{\pt}\te_{f(i,j)}})^{-1} , ~~~~ j> i .\eeq
Let $h_{i,j}=\frac{\te_{f(i,j)}' \rho_{M}^{\pt} \te_{f_{i,j}}}{\ati\atj}$. Since (\ref{eqntwo}) is true for all $j>i$,
\beq \label{eqnthree} t\le \min_{j>i} (1+\frac{1}{h_{i,j}})^{-1} .\eeq
For a given $\rho_{M}$, (\ref{eqnthree}) must be satisfied if $t \rho+ (1-t)\rho_{M} $ is separable. The largest $t$ for which there exists a $\rho_{M}\in\mbox{DM}(N)$ and where $t \rho+ (1-t)\rho_{M} $  is separable is upper bounded by the maximum of the right hand side (rhs) of (\ref{eqnthree}). That is,
\[ {\cal O}_{g}(\rho) \le \max_{\rho_{M}} \min_{j>i} (1+\frac{1}{h_{i,j}})^{-1} \]
or
\beq
\label{eqnfour} {\cal O}_{g}(\rho) \le (1+\frac{1}{\max_{\rho_{M}} \min_{j>i} h_{i,j}})^{-1} . \eeq
We will now consider the max-min problem  
\beq \label{eqnfive} \tt \dfn \max_{\rho_{M}} \min_{j>i} h_{i,j}. \eeq
Let $\rho_{M} = \sum_{i} \lambda_{i} e_{i} e_{i}' $ be a spectral decomposition of $\rho_{M}$, $||e_{i}||=1$, $e_{i}'e_{j}=0$, $i\ne j$, and where $||\cdot||$ denotes the L2 norm.  Denote the function $c(i,j)=m$ where $m$ is the indice where the vector $|i>\otimes|j>$ is equal to one, $|i>, |j>\in \com^{n}$. For example, if $n=2$, $|1>\otimes|2> = (0 1 0 0)'$, hence $c(1,2)=2$. Clearly then $c(i,j)=n(i-1)+j$.

Now rewrite the eigenvectors of $\rho_{M}$ as 
\[ e_{i} = \sum_{r,s} e_{i,c(r,s)} \ulr\otimes\uls . \]

Now,
\[ e_{i} e_{i}' = (\sum_{r,s} e_{i,c(r,s)} \ulr\otimes\uls) (\sum_{t,u} e_{i,c(t,u)} \ult\otimes \ulu)' \]
\[ \rho_{M} = \sum_{i} \lambda_{i} \sum_{r,s} \sum_{t,u} e_{i,c(r,s)} e_{i,c(t,u)}^{\ast} \ulr \ultp\otimes \uls \ulup \]
where $\ast$ denotes conjugate. The second partial transpose \cite{Q1:Pittenger2} of $\rho_{M}$ is
\beq \label{eqnsix}
\rho_{M}^{\pt} = \sum_{i} \lambda_{i}  \sum_{r,s} \sum_{t,u} e_{i,c(r,s)} e_{i,c(t,u)}^{\ast} \ulr \ultp\otimes \ulu \ulsp . \eeq
From Lemma $\ref{leig}$, we have $\te_{f(j,k)}' = \frac{1}{\sqrt{2}} (\uljp\otimes\ulkp - \uljp\otimes\ulkp)$,
so that 
\begin{eqnarray}
 \te_{f(j,k)}'\rho_{M}^{pt} \te_{f(j,k)} & =  &
\frac{1}{2}\sum_{i} \lambda_{i}  \sum_{r,s} \sum_{t,u} e_{i,c(r,s)} e_{i,c(t,u)}^{\ast} (<j\ulr<t\ulj\otimes<k\ulu <s\ulk-   \nonumber \\
 & & <j\ulr <t\ulk\otimes <k\ulu<s\ulj  -  <k\ulr<t\ulj\otimes<j\ulu<s\ulk  + \nonumber \\
& &  <k\ulr<t\ulk\otimes <j\ulu<s\ulj)  \nonumber \\
 & = & \frac{1}{2} \sum_{i} \lambda_{i} [e_{i,c(j,k)} e_{i,c(j,k)}^{\ast}  -  e_{i,c(j,j)} e_{i,c(k,k)}^{\ast}-e_{i,c(k,k)} e_{i,c(j,j)}^{\ast} +  e_{i,c(k,j)} e_{i,c(k,j)}^{\ast}] \nonumber \\
& = & \frac{1}{2} \sum_{i} \lambda_{i}[ |e_{i,c(j,k)}|^{2} + |e_{i,c(k,j)}|^{2} - 2\mbox{Re}(e_{i,c(j,j)} e_{i,c(k,k)}^{\ast})] ,\nonumber
\end{eqnarray}
where $\mbox{Re(x)}$ denotes the real part of $x$. Define
\[ g_{f(j,k)}^{(i)} \dfn |e_{i,c(j,k)}|^{2} + |e_{i,c(k,j)}|^{2} - 2\mbox{Re}(e_{i,c(j,j)} e_{i,c(k,k)}^{\ast}), \]
and the above becomes 
\[  \te_{f(j,k)}'\rho_{M}^{pt} \te_{f(j,k)} = \frac{1}{2} \sum_{i} \lambda_{i} g_{f(j,k)}^{(i)} . \]
Consider the matrix 
\beq
A = \left( \begin{array}{cccc} 
\frac{\lambda_{1}g_{1}^{(1)}}{2\tilde{a}_{1}\tilde{a}_{2}} & \frac{\lambda_{2}g_{1}^{(2)}}{2\tilde{a}_{1}\tilde{a}_{2}} & \cdots & \frac{\lambda_{N}g_{1}^{(1)}}{2\tilde{a}_{1}\tilde{a}_{2}} \\
\frac{\lambda_{1}g_{2}^{(1)}}{2\tilde{a}_{1}\tilde{a}_{3}} & \frac{\lambda_{2}g_{2}^{(2)}}{2\tilde{a}_{1}\tilde{a}_{3}} &\cdots & \frac{\lambda_{N}g_{2}^{(N)}}{2\tilde{a}_{1}\tilde{a}_{3}}  \\
\vdots & \ddots & & \vdots \\
\frac{\lambda_{1}g_{f(n-1,n}^{(1)}}{2\tilde{a}_{n-1}\tilde{a}_{n}} & \frac{\lambda_{2}g_{f(n-1,n)}^{(2)}}{2\tilde{a}_{n-1}\tilde{a}_{n}} & \cdots & \frac{\lambda_{N}g_{f(n-1,n)}^{(N)}}{2\tilde{a}_{n-1}\tilde{a}_{n}} 
\end{array} \right). 
\eeq
Then (\ref{eqnfive}) is identically
\beq \label{eqnseven} \tt = \max_{\rho_{M}}\min_{i} \sum_{j} A_{ij} . \eeq
Before the final step in the proof, consider 
\begin{eqnarray}
 \sum_{j=1}^{n(n-1)/2} g_{j}^{(i)} & = & \sum_{j<k} g_{f(j,k)}^{(i)} \nonumber \\ 
 & = & \sum_{j<k} |e_{i,c(j,k)}|^{2} + |e_{i,c(k,j)}|^{2} - e_{i,c(j,j)} e_{i,c(k,k)}^{\ast} -  e_{i,c(k,k)} e_{i,c(j,j)}^{\ast} \nonumber \\
 & = & \sum_{j\ne k} |e_{i,c(j,k)}|^{2} - e_{i,c(j,j)} e_{i,c(k,k)}^{\ast} \nonumber \\
 & = & (\sum_{j\ne k} |e_{i,c(j,k)}|^{2} + \sum_{j} |e_{i,c(j,j)}|^{2}) - (\sum_{j} |e_{i,c(j,j)}|^{2}+ \sum_{j\ne k}e_{i,c(j,j)} e_{i,c(k,k)}^{\ast}) \nonumber \\
 & = & 1 - (\sum_{j} |e_{i,c(j,j)}|^{2}+ \sum_{j\ne k}e_{i,c(j,j)} e_{i,c(k,k)}^{\ast}) \nonumber \\
 & = & 1 - | \sum_{j} e_{i,c(j,j)}|^{2} \nonumber \\
 \label{eqneight} & \le & 1 .
\end{eqnarray}
Now, (\ref{eqnseven}) can be written as
\beq \label{eqnnine} 
\tt = \max_{\rho_{M}}\min_{j<k} \sum_{i} \frac{\lambda_{i}g_{f(j,k)}^{(i)}}{2\tilde{a}_{j}\tilde{a}_{k}} .
\eeq
Let $\alpha_{j,k} = \sum_{i} \lambda_{i} g^{(i)}_{f(j,k)}.$ Then (\ref{eqnnine}) becomes
\[ \tt = \max_{\rho_{M}}\min_{j<k} \frac{\alpha_{j,k}}{2\tilde{a}_{j}\tilde{a}_{k}} . \]
Summing over $\alpha_{j,k}$ with $j<k$ we have \begin{eqnarray*}
\sum_{j<k} \alpha_{j,k} & = & \sum_{j<k} \sum_{i} \lambda_{i} g^{(i)}_f(j,k) \\
& = & \sum_{i} \lambda_{i} \sum_{j<k} g^{(i)}_{f(j,k)} \\
& & \mbox{and from (\ref{eqneight})} \\
& \le & \sum_{i} \lambda_{i} \\
& = & 1.
\end{eqnarray*}
Hence $\tt$ in (\ref{eqnnine}) can be upper bounded by 
\begin{eqnarray*}  \tt \le \max_{\rho_{M}} \min_{i} \delta_{i}\beta_{i} 
~~\mbox{where}~\beta_{f(j,k)} & = & \frac{1}{2\tilde{a}_{j}\tilde{a}_{k}}, ~j<k\\
~~~~~~~~~\delta_{f(j,k)} & = & \alpha_{j,k}, ~j<k, \sum_{i} \delta_{i} \le 1
\end{eqnarray*} 
or
\[  \tt \le \max_{\delta_{i}\le 1} \min_{i} \delta_{i}\beta_{i} ~~~\mbox{where}~~~ \beta_{f(j,k)}  =  \frac{1}{2\tilde{a}_{j}\tilde{a}_{k}}, ~j<k \]
and rewritten as
\beq \label{eqntwelve} \tt \le \max_{\begin{array}{c}y\le \delta_{i}\beta_{i} \\ \sum_{i} \delta_{i}\le 1 \end{array}} y .\eeq
Consider the candidate solution  to (\ref{eqntwelve}) of $\delta_{i} = \delta_{i}^{(1)} = \frac{K}{\beta_{i}}$ where $K$ is a constant.  
Since $\sum_{i} \delta_{i}^{(1)} \le 1$, it follows that $K \le \frac{1}{\sum_{j<k} 2\tilde{a}_{j}\tilde{a}_{k}}$. Note that the denominator is equal to $(\sum_{i} \tilde{a}_{i})^{2}-1$ which is equal to $\mbox{R}_{s}(\rho)$ from Eqn. (\ref{eqnrob}). Hence
\beq \label{eqnthirteen} K\le \frac{1}{\mbox{R}_{s}(\rho)} . \eeq
For this choice of $\delta_{i}^{(1)}$, the rhs of Eqn. (\ref{eqntwelve}) is maximized when $K$ is on the boundary in Eqn. (\ref{eqnthirteen}) i.e. $K=\frac{1}{\mbox{R}_{s}(\rho)}$ and for which $\sum_{i} \delta_{i}^{(1)}=1$. Then 
$\delta_{i}^{(1)} = \frac{1}{\mbox{R}_{s}(\rho)\beta_{i}}$, and 
\[  \max_{\begin{array}{c}y\le \delta_{i}\beta_{i} \\ \sum_{i} \delta_{i}\le 1 \end{array}} \ge \max_{\begin{array}{c}y\le \delta_{i}^{(1)}\beta_{i} \\ \sum_{i} \delta_{i}^{(1)} = 1 \end{array}} = \frac{1}{\mbox{R}_{s}(\rho)} . \]
We will now show that no other larger solutions to (\ref{eqntwelve}) exist. Suppose that such a solution $y_{2}$ exists and let $\delta_{i}=\delta_{i}^{(2)}$ be the associated parameters in (\ref{eqntwelve}).  Since $y_{2}>\frac{1}{\mbox{R}_{s}(\rho)}$, there must exist a $\delta_{e}>0$ whereby $y_{2} = \frac{1}{\mbox{R}_{s}(\rho)}+ \delta_{e}$. From (\ref{eqntwelve}) it follows that
\begin{eqnarray*}
y_{2} & \le & \delta_{i}^{(2)} \beta_{i} ~~~~~~~\forall i \\
\frac{1}{\mbox{R}_{s}(\rho)} + \delta_{e} & \le & \delta_{i}^{(2)} \beta_{i} ~~~~~~~\forall i \\
\delta_{i}^{(1)}\beta_{i} + \delta_{e} & \le & \delta_{i}^{(2)} \beta_{i} ~~~~~~~\forall i \\
0 < \delta_{e} & \le & (\delta_{i}^{(2)} - \delta_{i}^{(1)})\beta_{i} ~~~~~~~\forall i \\
& &\mbox{from which it follows that} \\
\delta_{i}^{(2)} & > & \delta_{i}^{(1)} ~~~~~~\forall i . \\
& & \mbox{Summing both sides}  \\
\sum_{i} \delta_{i}^{(2)} & > & \sum_{i} \delta_{i}^{(1)} \\
& & \mbox{which implies} \\
\sum_{i} \delta_{i}^{(2)} & > & 1 .
\end{eqnarray*}
This contradicts the assumption in (\ref{eqntwelve}) that $\sum_{i} \delta_{i}^{(2)} > 1 $. Hence
\[   \tt \le \frac{1}{\mbox{R}_{s}(\rho)} , \]
and so we have upper bounded the rhs of (\ref{eqnfour}) to arrive at
\[ {\cal O}_{g}(\rho) \le \frac{1}{1+\mbox{R}_{s}(\rho)} . \]
Since from (\ref{eqngenandvt}) we know that ${\cal O}_{g}(\rho) \ge  {\cal O}_{s}(\rho)$, the result follows.  \mybox

An interesting question is whether or not an optimal $\rho_{M}$ in Eqn. (\ref{eqndef}) (optimal in the sense of maximizing ${\cal O}_{g}$), is necessarily separable.  We already know that such separable states $\rho_{M}$ can be constructed via the construction given in \cite{Q1:Vidal2}. It is shown by counterexample that one can also construct $\rho_{M}$ that are entangled.  To construct the counterexample, consider an entangled pure state $\rho$ with $a={\cal O}_{g}(\rho) $.  Assume that $\rho$ is in the simplified form $\rho=\psi\psi'$, with $\psi=\sum_{i} \tilde{a_{i}} |i> \otimes |i>$ where $\tilde{a_{i}}\ge 0$ are the Schmidt coefficients of $\psi$ and $|i>$ is in the natural basis. Let $G=-a\rho/(1-a)$ and note that all the elements in $G$ satisfy $G_{i,j}\le 0$, where $G_{i,j}$ represents the $i$th column and $j$th row of $G$. Let $G^{(2)}$ be the matrix $G$ except with the diagonal elements removed.   Now $G^{(2)}$ is generally not positive definite so we will replace the diagonal elements with values from the Gersgoren disks \cite[p. 344]{Q1:Horn}. That is, $G^{(2)}_{i,i}= - \sum_{k} G_{i,k}$. This will guarantee that $G^{(2)}$ is positive semidefinite. Now, $a\rho+(1-a)G^{(2)}$ is clearly a diagonal matrix. It can also be seen that the diagonal elements are positive and can be verified the sum of the diagonal elements is 1 and $G^{(2)}$ is typically entangled. Hence $a\rho+(1-a)G^{(2)}$ is separable and there exists entangled matrices $\rho_{M}$ that optimize the generalized robustness of entanglement. Additionally, note that if there are two optimal solutions for $\rho_{M}$ in (\ref{eqndef}), call them $\rho_{M_{1}}$ and $\rho_{M_{2}}$, then it is easily proven that any convex combination, i.e. $t\rho_{M_{1}}+(1-t) \rho_{M_{2}}$, $0\le t\le 1$, also is an optimal solution  for $\rho_{M}$ in (\ref{eqndef}).  Since we have seen that both entangled and non-entangled solutions exist, this further implies that there can be an infinite number of solutions for $\rho_{M}$.

\section{Conclusions}
Vidal and Tarrach considered robustness by considering how much one can mix a state with an arbitrary separable state such that the combination is separable. This result has been extended by allowing any state to mix with the state such that the combination is separable.  It was found that the state is just as robust as before.  This is somewhat a surprising result since the volume of state space becomes dominated by the entangled states in large dimension as seen in \cite{Q1:Zyczkowski}. The author initially expected that the robustness as measured by $\mbox{R}_{g}(\rho)$ would decrease as compared to $\mbox{R}_{s}(\rho)$.  However, it was found that there is no degradation.  This result shows that entangled states that are generally robust (in terms of this definition) when only non-entangled states are present would be expected to robust in similar situations where there are other entangled states present that can mix with the state to cause disentanglement. Extensions to this work would be to determine the generalized robustness of entanglement for mixed states. 

\vspace{.25in}

{\normalsize \noindent{\bf Acknowledgements:} The author acknowledges helpful discussions with R. Lockhart, M. Rubin, and A. Pittenger and the Office of Naval Research for the support of this work.}

\vspace{.25in}

{\it Note added in proof:} After this work was submitted, the author learned of two recent relevant works \cite{Q1:Harrow}
and \cite{Q1:Verstraete}.  The main result on the generalized robustness of entanglement for pure states found in this paper was also found independently by A. Harrow and M. Nielsen and derived in a rather nice compact manner \cite{Q1:Harrow}. The authors also extend the notion of state robustness for the application of quantum logic gate robustness.  In \cite{Q1:Verstraete}, F. Verstraete and H. Verschelde examined the problem of the maximal achievable fidelity optimized over all local operations and classical communication (LOCC) operations. They found an interesting equivalence between this latter problem and the generalized robustness of entanglement for the case of two qubits. They also developed a solution for the case of two qubits in a mixed state, which reduces to the solution in this paper when the two qubits are pure.

\section{\protect\normalsize APPENDIX}

Proofs of Theorem 1 and 2. These proofs follow from \cite{Q1:Vidal2} with minor modification.

\newtheorem{thm}{Theorem}
\begin{thm}
$\rob$ is convex, i.e. \[ \mbox{R}(p \rho_{1}+ (1-p) \rho_{2}) \le p \mbox{R}(\rho_{1})+ (1-p) \mbox{R}(\rho_{2}) .\]
\end{thm}

\noindent{Proof}:
Define $t\dfn p \mbox{R}(\rho_{1})+ (1-p) \mbox{R}(\rho_{2})$ and $\rho\dfn p\rho_{1}+(1-p)\rho_{2}$. From the definition of robustness, there exists $\rho_{M,1}\subset\dm(N)$ and $\rho_{M,2}\subset\dm(N)$, where $\dm(N)$ represents the set of $N$ by $N$ density matrices, such that \[ \rho_{1} = (1+\rb(\rho_{1}))\rho_{s,1} - \rb(\rho_{1}) \rho_{M,1} \] 
\[ \rho_{2} = (1+\rb(\rho_{2}))\rho_{s,2} - \rb(\rho_{1}) \rho_{M,2} \] 
and for which $\rho_{s,1},\rho_{s,2}\in\sep$. Now, note that the matrix $\rho_{s}$ as defined by
\[ \rho_{s} \dfn  \frac{1}{(1+t)}(p(1+ \rb(\rho_{1})\rho_{s,1} +(1-p)(1+\rb(\rho_{2}))\rho_{s,2}) \]
is separable. It can be shown that
\[ \rho = (1+t) \rho_{s} - t \rho_{M,3} \]
where $\rho_{M,3} \dfn \frac{1}{t}(p \rb(\rho_{1})\rho_{s,1}+(1-p)\rb(\rho_{2})\rho_{s,2}) .$
Hence $\rb(\rho)\le t$. \mybox

\begin{thm}
\label{thmtwo2}
$\rob = \mbox{R}_{g}(U_{L} \rho U_{L}) $, where $U_{L}$ is a local unitary transformation of the form $U_{L}= U_{1} \otimes U_{2}$.
\end{thm}

\noindent{Proof}:
By definition, there exists a decomposition $\rho=(1+\rb(\rho))\rho_{s} - \rb(\rho) \rho_{M}$ with $\rho_{s}\in\sep$. Then
\[ U_{L} \rho U_{L}\ctr = (1+\rb(\rho))U_{L} \rho_{s}U_{L}\ctr-\rb(\rho) U_{L}\rho_{M}U_{L}\ctr . \]
Since $U_{L}\rho U_{L}\ctr\in \sep$ then $\rb(U_{L}\rho U_{L}\ctr)\le \rb(\rho).$
On the other hand if one considers $\rho_{2}\dfn U_{L}\rho U_{L}\ctr$, there exists by definition a decomposition $\rho_{2} = (1+\rb(\rho_{2}))\rho_{s,2}-\rb(\rho_{2})\rho_{M,2}.$ Multiplying both sides by $U_{L}\ctr$ and $U_{L}$, gives
\[ \rho = (1+\rb(\rho_{2}))U_{L}\ctr\rho_{s,2}U_{L} - \rb(\rho_{2})U_{L}\ctr\rho_{M,2}U_{L} \]
which implies $\rb(\rho)\le \rb(\rho_{2}) .$ \mybox

%\bibliographystyle{h-physrev}
%\bibliography{C:/mike/texbib/q1}

\begin{thebibliography}{1}

\bibitem{Q1:Vidal2}
G.~Vidal and R.~Tarrach,
\newblock Phys. Rev. A {\bf 59}, 141, (1999).

\bibitem{Q1:Zeh2}
H.~Zeh,
\newblock Decoherence: Basic concepts and their interpretation, 1995,
\newblock quant-ph/9506020.

\bibitem{Q1:Sorensen}
A.~S. Sorensen,
\newblock Phys. Rev. A {\bf 65}, 043610, (2002).

\bibitem{Q1:Zyczkowski}
K.~Zyczkowski, P.~Horodecki, A.~Sanpera, and M.~Lewenstein,
\newblock Phys. Rev. A {\bf 58}, 883 (1998).

\bibitem{Q1:Steiner4}
R.~B. Lockhart and M.~J. Steiner,
\newblock Phys. Rev. A {\bf 65}, 022107 (2002).

\bibitem{Q1:Peres}
A.~Peres,
\newblock Phys. Rev. Lett. {\bf 77}, 1413 (1996).

\bibitem{Q1:Pittenger2}
A.~O. Pittenger and M.~H. Rubin,
\newblock Convexity and the separability problem of quantum mechanical density
  matrices, 2001, quant-ph/0103038.

\bibitem{Q1:Horn}
R.~A. Horn and C.~R. Johnson,
\newblock {\em Matrix Analysis} (Cambridge University Press, 1985).

\bibitem{Q1:Harrow}
A. Harrow and M. Nielsen, eprint quant-ph/0301108 (2003)

\bibitem{Q1:Verstraete}
F. Verstraete and H. Verschelde, Phys. Rev. Lett. 90, 097901 (2003)

\end{thebibliography}
\end{document}